\documentclass[a4paper]{jpconf}
\usepackage{graphicx}
\usepackage{amsmath,amsbsy,amsfonts,mathrsfs,bm}

\begin{document}

\title{Supersymmetry and non-Abelian geometric phase for a free particle on a circle with point-like interactions}
\author{Satoshi Ohya$^{1,2}$}
\address{$^1$ Department of Physics, Faculty of Nuclear Sciences and Physical Engineering, Czech Technical University in Prague, Pohrani\v{c}n\'{i} 1288/1, 40501 D\v{e}\v{c}\'{i}n, Czech Republic}
\address{$^2$ Doppler Institute for Mathematical Physics and Applied Mathematics, Czech Technical University in Prague, B\v{r}ehov\'{a} 7, 11519 Prague, Czech Republic}
\ead{ohyasato@fjfi.cvut.cz}

\begin{abstract}
Though not so widely appreciated in the literature, supersymmetric quantum mechanics provides an ideal playground for studying non-Abelian geometric phase, because supersymmetry always guarantees degeneracies in energy levels.
In this paper we first present a simple supersymmetric model for a free particle on a circle with point-like interactions that exhibits $\mathscr{N} = 2$ supersymmetry and doubly degenerate energy levels.
We then show that Berry's connection in this model is given by the Wu-Yang-like magnetic monopole in $SU(2)$ Yang-Mills gauge theory.
This article is largely based on our recent work \cite{Ohya:2014ska}.
\end{abstract}

%-----------------------------------------------------------------------------------------------
% SECTION 1
%-----------------------------------------------------------------------------------------------
\section{Introduction}
Berry's geometric phase \cite{Berry:1984jv} and its non-Abelian generalization \cite{Wilczek:1984dh} are ubiquitous phenomena in quantum physics when model parameters are adiabatically driven along a closed loop on the parameter space.
They appear in a wide spectrum of physics and have applications in many disciplines, ranging from condensed matter physics to nuclear and high energy physics.

The purpose of this paper is to introduce a new example of quantum mechanical models that exhibits nontrivial non-Abelian geometric phase.
In order to realize non-Abelian geometric phase, we must at least have:
\begin{enumerate}
\item degenerate energy levels; and
\item a certain number of (experimentally controllable) parameters.
\end{enumerate}
Let us first look at the criterion (i).
As is well-known, degeneracies in energy levels are closely tied to symmetries of the systems.
A typical example of such symmetries is time-reversal invariance, which results in the Kramers degeneracy in Fermi systems and plays an important role in geometric phase and/or topological invariants in molecular systems \cite{Mead:1987gk,Avron:1988ty,Avron:1989}.
In this paper we focus on yet another symmetry which induces spectral degeneracy: that is, \textit{supersymmetry}, which also guarantees degenerate energy levels between ``bosonic'' and ``fermionic'' degrees of freedom of the system.\footnote{This is always true except for the ground state(s): In supersymmetric quantum mechanics the ground state degeneracy is related to the topological index $\Tr_{\mathcal{H}}(-1)^{F}$ rather than symmetries \cite{Witten:1982df}. In \cite{Pedder:2007ff,Sonner:2008be} non-Abelian geometric phase over the space of degenerate ground states has been studied in the context of supersymmetric quantum mechanics.}

Let us next look at the criterion (ii).
A typical example of (experimentally controllable) parameters discussed in the context of geometric phase is electromagnetic field.
Here we would like to point out that generic point-like interactions may also serve the purpose, because point-like interactions are all known to be described by $U(2)$ family of boundary conditions \cite{Fulop:1999pf} and characterized by $\dim U(2) = 4$ independent continuous parameters.
Though, as far as we know, $U(2)$ family of point-like interactions have not yet realized in laboratory experiments, they are completely allowed in quantum mechanics and hence will provide a promising ingredient of future nanotechnology.

The aim of the paper is to present a canonical example of quantum mechanical models with point-like interactions that exhibits supersymmetry, spectral degeneracy, and nontrivial non-Abelian geometric phase.\footnote{In \cite{Exner:1999,Cheon:2000tq} Abelian geometric phase has been studied in simple models of point-like interactions.}
As in any study of theoretical physics, it is tempting to construct a minimal model that captures all the essence of the ideas.
In this paper we consider a single spinless particle on a circle of circumference $2\ell$ that freely propagates in the bulk yet interacts only at the origin $x = 0$ and its antipodal point $x = \ell$, where $x = 0$ and $x = 2\ell$ are identified.
The bulk dynamics is governed by the Schr\"{o}dinger equation for a free particle
\begin{equation}
-\frac{\hbar^{2}}{2m}\frac{d^{2}}{dx^{2}}\psi(x) = E\psi(x), \qquad x \neq 0, \ell. \label{eq:1}
\end{equation}
Point-like interactions, on the other hand, are all described by the boundary conditions consistent with \textit{Kirchhoff's law} of probability current at $x = 0$ and $\ell$ \cite{Fulop:1999pf}:
\begin{equation}
j(0_{+}) = j(2\ell_{-})
\qquad\text{and}\qquad
j(\ell_{-}) = j(\ell_{+}), \label{eq:2}
\end{equation}
where $j(x) = \frac{\hbar}{2im}(\psi^{\ast}(x)\psi^{\prime}(x) - \psi^{\prime\ast}(x)\psi(x))$ is the local probability current (with prime (${}^{\prime}$) indicating the derivative with respect to $x$) and $j(a_{\pm})$ stand for the limits approaching $x = a$ from above and from below.
The conditions \eqref{eq:2} are quadratic in wavefunctions, but they can be linearized and are known to enjoy the following $U(2)$ family of solutions at each point \cite{Fulop:1999pf}:
\begin{subequations}
\begin{align}
(1_{2} - U)
\begin{pmatrix}
\psi(0_{+})\\
\psi(2\ell_{-})
\end{pmatrix}
- iL_{0}(1_{2} + U)
\begin{pmatrix}
\psi^{\prime}(0_{+})\\
-\psi^{\prime}(2\ell_{-})
\end{pmatrix}
= 0, \qquad U \in U(2), \label{eq:3a}\\
(1_{2} - \Tilde{U})
\begin{pmatrix}
\psi(\ell_{-})\\
\psi(\ell_{+})
\end{pmatrix}
+ iL_{0}(1_{2} + \Tilde{U})
\begin{pmatrix}
\psi^{\prime}(\ell_{-})\\
-\psi^{\prime}(\ell_{+})
\end{pmatrix}
= 0, \qquad \Tilde{U} \in U(2), \label{eq:3b}
\end{align}
\end{subequations}
where $1_{2}$ stands for the $2 \times 2$ unit matrix and $L_{0}$ is an arbitrary length scale that needs to be introduced on account of dimensional analysis.
Equations \eqref{eq:3a} and \eqref{eq:3b} describe the most general point-like interactions in quantum mechanics consistent with Kirchhoff's law of probability current \eqref{eq:2}, or, equivalently, self-adjointness of the free Hamiltonian.
The full parameter space of the model is thus $U(2) \times U(2)$.
Though energy levels are not degenerate in general, there indeed exists a subspace $\mathcal{M}_{\text{SUSY}} \subset U(2) \times U(2)$ in which $\mathscr{N} = 2$ supersymmetry\footnote{Here $\mathscr{N}$ refers to the number of real (i.e. self-adjoint) supercharges.} emerges and energy levels become doubly degenerate (up to the question of ground state degeneracy).
As shown in \cite{Ohya:2014ska}, the supersymmetric subspace $\mathcal{M}_{\text{SUSY}}$ is given by the following direct product:
\begin{equation}
\mathcal{M}_{\text{SUSY}}
= 	U(1) \times \frac{U(2)}{U(1) \times U(1)}
\cong S^{1} \times S^{2}. \label{eq:4}
\end{equation}
It can be shown that only the first factor controls the energy eigenvalues and the second factor does not affect the energy spectrum at all: the coset factor $U(2)/(U(1) \times U(1))$ describes the isospectral family of the system and hence provides an ideal playground for studying geometric phase.
The goal of the paper is to report that the Berry connection on $\mathcal{M}_{\text{SUSY}}$ with fixed $U(1)$ parameter is given by a magnetic monopole-like configuration in non-Abelian gauge theory.

The organization of the paper is as follows.
In section \ref{sec:2} we construct $\mathscr{N} = 2$ supersymmetric quantum mechanics in this system and solve the energy eigenvalue problems.
We then compute non-Abelian Berry's connection on the parameter space $U(2)/(U(1) \times U(1)) \cong S^{2}$ in section \ref{sec:3} and show that it is nothing but the Wu-Yang-like magnetic monopole in $SU(2)$ Yang-Mills gauge theory.
Section \ref{sec:4} is devoted to conclusions and discussions.

In the rest of the paper we will work in the units $\hbar = 2m = 1$.

%-----------------------------------------------------------------------------------------------
% SECTION 2
%-----------------------------------------------------------------------------------------------
\section{Supersymmetric point-like interactions} \label{sec:2}
The key to construct $\mathscr{N} = 2$ supersymmetric quantum mechanics for a free particle on a circle with two point-like interactions is to find the generic form of fermion parity operator $(-1)^{F}$, which should satisfy $((-1)^{F})^{2} = 1$ and have two eigenvalues $+1$ and $-1$.
In this section we first introduce $(-1)^{F}$ in our model and then discuss point-like interaction invariant under supersymmetry transformations.
We also explicitly construct $\mathscr{N} = 2$ supersymmetry algebra and solve the superspectrum in this model.

%-----------------------------------------------------------------------------------------------
% SECTION 2.1
%-----------------------------------------------------------------------------------------------
\subsection{Folding trick}
In the following discussions it is convenient to work in the so-called \textit{folding picture},\footnote{Folding trick has been developed in the context of boundary conform field theory \cite{Oshikawa:1996dj,Bachas:2001vj}.} which is just a change of viewpoint   from ``scalar'' quantum mechanics (i.e. quantum mechanics with scalar-valued wavefunction) on a circle of circumference $2\ell$ to ``vector'' quantum mechanics (i.e. quantum mechanics with vector-valued wavefunction) on an interval of length $\ell$.
To do this, let us consider the wavefunction on the upper- and lower-semicircles separately and embed them into a single two-component vector-valued function as follows:
\begin{align}
\bm{\Psi}(x)
:= 	\begin{pmatrix}
	\psi(x) \\
	\psi(2\ell - x)
	\end{pmatrix},
	\qquad 0 < x < \ell. \label{eq:5}
\end{align}
(Throughout the paper we will use boldface symbols for vectors.)
The boundary conditions \eqref{eq:3a} and \eqref{eq:3b} are then cast into the following forms:
\begin{subequations}
\begin{align}
(1_{2} - U)\bm{\Psi}(0) - iL_{0}(1_{2} + U)\bm{\Psi}^{\prime}(0)
&= 	0, \label{eq:6a}\\
(1_{2} - \Tilde{U})\bm{\Psi}(\ell) + iL_{0}(1_{2} + \Tilde{U})\bm{\Psi}^{\prime}(\ell)
&= 	0. \label{eq:6b}
\end{align}
\end{subequations}
Here and hereafter we simply write $\bm{\Psi}(0)$ and $\bm{\Psi}(\ell)$ for $\bm{\Psi}(0_{+})$ and $\bm{\Psi}(\ell_{-})$.
In the folding picture the Hilbert space is $\mathcal{H} = L^{2}(0,\ell) \otimes \mathbb{C}^{2}$ and the free Hamiltonian that acts on $\mathcal{H}$ is given by the $2 \times 2$ matrix-valued operator, $H = \left(\begin{smallmatrix}h & 0 \\ 0 & h\end{smallmatrix}\right)$, where $h = -d^{2}/dx^{2}$.
The boundary conditions \eqref{eq:6a} and \eqref{eq:6b} specify the most general self-adjoint domain of the matrix-valued operator $H$.

Let us next introduce the fermion parity $(-1)^{F}$ in this model.
As discussed in \cite{Nagasawa:2002un,Ohya:2014ska}, $(-1)^{F}$ is given by the following hermitian unitary operator:
\begin{align}
\mathcal{Z}: \bm{\Psi}(x) \mapsto (\mathcal{Z}\bm{\Psi})(x) := Z\bm{\Psi}(x), \label{eq:7}
\end{align}
where $Z$ is a generic $2 \times 2$ hermitian unitary matrix satisfying $Z = Z^{\dagger} = Z^{-1}$, or $Z^{2} = 1_{2}$.
Such hermitian unitary matrix is parameterized as $Z = \bm{n} \cdot \bm{\sigma}$, where $\bm{\sigma} = (\sigma_{1}, \sigma_{2}, \sigma_{3})$ is the vector of Pauli matrices and $\bm{n} = (n_{1}, n_{2}, n_{3})$ is a real unit 3-vector fluffing the condition $n_{1}^{2} + n_{2}^{2} + n_{3}^{2} = 1$.
The parameter space of $Z$ is therefore 2-sphere $S^{2}$, which turns out to provide the coset factor in the supersymmetric parameter space \eqref{eq:4}.
Notice that the unitary transformation \eqref{eq:7} commutes with the Hamiltonian and leaves the Schr\"{o}dinger equation unchanged.
However, $\mathcal{Z}$ does not leave the boundary conditions \eqref{eq:6a} and \eqref{eq:6b} unchanged in general and hence is not a symmetry for generic point-like interactions.
It is easy to show that the boundary conditions remain invariant under $\mathcal{Z}$ if and only if both $U$ and $\Tilde{U}$ commute with $Z$, $[U, Z] = [\Tilde{U}, Z] = 0$ \cite{Fulop:2003,Ohya:2014ska}.
The solutions to these conditions are parameterized as follows:
\begin{subequations}
\begin{align}
U
&= 	\mathrm{e}^{i\alpha_{+}}P_{+} + \mathrm{e}^{i\alpha_{-}}P_{-}, \label{eq:8a}\\
\Tilde{U}
&= 	\mathrm{e}^{i\Tilde{\alpha}_{+}}P_{+} + \mathrm{e}^{i\Tilde{\alpha}_{-}}P_{-}, \label{eq:8b}
\end{align}
\end{subequations}
where $P_{\pm} = (1_{2} \pm Z)/2$ are projection operators and $\{\alpha_{+}, \alpha_{-}, \Tilde{\alpha}_{+}, \Tilde{\alpha}_{-}\}$ are eigenphases of $U$ and $\Tilde{U}$.
The fermion parity $(-1)^{F} = \mathcal{Z}$ becomes the symmetry of the system if and only if $U$ and $\Tilde{U}$ are given by \eqref{eq:8a} and \eqref{eq:8b}.
In this case $(-1)^{F} = \mathcal{Z}$ provides a good grading operator such that the Hilbert space splits into two orthogonal subspaces, $\mathcal{H} = \mathcal{H}_{+} \oplus \mathcal{H}_{-}$, where $\mathcal{H}_{\pm} = \{\bm{\Psi} \in \mathcal{H} \mid \mathcal{Z}\bm{\Psi} = \pm\bm{\Psi}\}$ are ``bosonic'' and ``fermionic'' sectors of the model.
Correspondingly, any element $\bm{\Psi} \in \mathcal{H}$ of the Hilbert space $\mathcal{H}$ is decomposed as follows:
\begin{align}
\bm{\Psi}(x)
= 	\bm{\Psi}_{+}(x) + \bm{\Psi}_{-}(x), \label{eq:9}
\end{align}
where
\begin{align}
\bm{\Psi}_{\pm}(x)
= 	P_{\pm}\bm{\Psi}(x)
= 	\Psi_{\pm}(x)\bm{e}_{\pm}. \label{eq:10}
\end{align}
Here $\bm{e}_{\pm}$ are orthonormal eigenvectors of $Z$ that satisfy the eigenvalue equations $Z\bm{e}_{\pm} = \pm\bm{e}_{\pm}$, the orthonormality $\bm{e}_{\alpha}^{\dagger}\bm{e}_{\beta} = \delta_{\alpha\beta}$ and the completeness $\bm{e}_{+}\bm{e}_{+}^{\dagger} + \bm{e}_{-}\bm{e}_{-}^{\dagger} = 1_{2}$.
Notice that the projection operators can also be written as $P_{\pm} = \bm{e}_{\pm}\bm{e}_{\pm}^{\dagger}$.
In what follows we will call the set of eigenvectors $\{\bm{e}_{+}, \bm{e}_{-}\}$ the basis and $\Psi_{\pm} = \bm{e}_{\pm}^{\dagger}\bm{\Psi}$ the components.
It is easy to check that the $\mathcal{Z}$-invariant boundary conditions completely decouple into the ``bosonic'' and ``fermionic'' parts and the components $\Psi_{\pm}$ satisfy the following Robin boundary conditions:
\begin{subequations}
\begin{align}
\sin\left(\frac{\alpha_{\pm}}{2}\right)\Psi_{\pm}(0)
+ L_{0}\cos\left(\frac{\alpha_{\pm}}{2}\right)\Psi_{\pm}^{\prime}(0)
&= 	0, \label{eq:11a}\\
\sin\left(\frac{\Tilde{\alpha}_{\pm}}{2}\right)\Psi_{\pm}(\ell)
- L_{0}\cos\left(\frac{\Tilde{\alpha}_{\pm}}{2}\right)\Psi_{\pm}^{\prime}(\ell)
&= 	0. \label{eq:11b}
\end{align}
\end{subequations}

%-----------------------------------------------------------------------------------------------
% SECTION 2.2
%-----------------------------------------------------------------------------------------------
\subsection{Supersymmetric boundary conditions}
Let us next introduce supersymmetry transformations.
To this end, let us first introduce the following one-parameter family of first-order differential operators:
\begin{align}
A_{\alpha}^{\pm}
= 	\pm\frac{d}{dx} + \frac{1}{L(\alpha)},  \label{eq:12}
\end{align}
where $L(\alpha) := L_{0}\cot(\alpha/2)$.
An important observation is that the free Hamiltonian $h = -d^{2}/dx^{2}$ is factorized as $h = A_{\alpha}^{\mp}A_{\alpha}^{\pm} - 1/L(\alpha)^{2}$ such that the Schr\"{o}dinger equations $h\Psi_{\pm} = E\Psi_{\pm}$ can be written as $A_{\alpha}^{-}A_{\alpha}^{+}\Psi_{+} = (E + 1/L(\alpha)^{2})\Psi_{+}$ and $A_{\alpha}^{+}A_{\alpha}^{-}\Psi_{-} = (E + 1/L(\alpha)^{2})\Psi_{-}$, which imply the following relations:
\begin{subequations}
\begin{align}
A_{\alpha}^{+}\Psi_{+}(x)
&= 	\sqrt{E + \frac{1}{L(\alpha)^{2}}}\Psi_{-}(x), \label{eq:13a}\\
A_{\alpha}^{-}\Psi_{-}(x)
&= 	\sqrt{E + \frac{1}{L(\alpha)^{2}}}\Psi_{+}(x). \label{eq:13b}
\end{align}
\end{subequations}
These equations give the supersymmetry transformations that map the ``bosonic'' state $\Psi_{+}$ to the ``fermionic'' state $\Psi_{-}$ and vice versa.
It should be emphasized that the Robin boundary conditions \eqref{eq:11a} and \eqref{eq:11b} are not invariant under these transformations for generic values of $\alpha_{\pm}$ and $\Tilde{\alpha}_{\pm}$.
The boundary conditions invariant under the supersymmetry transformations are classified in \cite{Ohya:2014ska}, and it is shown that there are four distinct supersymmetric boundary conditions, which are summarized in table \ref{tab:1}.
We note that the supersymmetric boundary conditions admit only one parameter $\alpha$, which is an angle parameter and provides the $U(1)$ factor in \eqref{eq:4}.
\begin{table}
\caption{\label{tab:1}Supersymmetric boundary conditions.}
\begin{center}
\begin{tabular}{lll}
\br
 		& ``bosonic'' sector 	& ``fermionic'' sector \\
\mr
type DD 	& $\Psi_{+}(0) = 0 = \Psi_{+}(\ell)$ 	& $(A^{-}_{\alpha}\Psi_{-})(0) = 0 = (A^{-}_{\alpha}\Psi_{-})(\ell)$ \\
type RR 	& $(A^{+}_{\alpha}\Psi_{+})(0) = 0 = (A^{+}_{\alpha}\Psi_{+})(\ell)$ 	& $\Psi_{-}(0) = 0 = \Psi_{+}(0)$ \\
type DR 	& $\Psi_{+}(0) = 0 = (A^{+}_{\alpha}\Psi_{+})(\ell)$ 	& $(A^{-}_{\alpha}\Psi_{-})(0) = 0 = \Psi_{-}(\ell)$ \\
type RD 	& $(A^{+}_{\alpha}\Psi_{+})(0) = 0 = \Psi_{+}(\ell)$ 	& $\Psi_{-}(0) = 0 = (A^{-}_{\alpha}\Psi_{-})(\ell)$ \\
\br
\end{tabular}
\end{center}
\end{table}

%-----------------------------------------------------------------------------------------------
% SECTION 2.3
%-----------------------------------------------------------------------------------------------
\subsection{$\mathscr{N}=2$ supersymmetry algebra}
Now we are in a position to introduce $\mathscr{N} = 2$ supersymmetry algebra that consists of a single self-adjoint Hamiltonian $H$, two self-adjoint supercharges $Q_{1}$ and $Q_{2}$, and a single fermion parity $(-1)^{F}$.
For the following discussions it is convenient to consider the nilpotent supercharges $Q^{\pm}$ which are given by the linear combinations $Q^{\pm} = (Q_{1} \pm iQ_{2})/2$.
Let us first work in the basis in which the fermion parity becomes diagonal.
In that basis the wavefunction becomes $\bm{\Psi} = (\Psi_{+}, \Psi_{-})^{T}$ such that the operators $H$, $(-1)^{F}$ and $Q^{\pm}$ are given by the following standard forms:
\begin{subequations}
\begin{align}
H
&= 	\begin{pmatrix}
	h 	& 0 \\
	0 	& h
	\end{pmatrix}, \label{eq:14a}\\
(-1)^{F}
&= 	\begin{pmatrix}
	1 	& 0 \\
	0 	& -1
	\end{pmatrix}, \label{eq:14b}\\
Q^{+}
&= 	\begin{pmatrix}
	0 			& 0 \\
	A^{+}_{\alpha} 	& 0
	\end{pmatrix}, \label{eq:14c}\\
Q^{-}
&= 	\begin{pmatrix}
	0 	& A^{-}_{\alpha} \\
	0 	& 0
	\end{pmatrix}. \label{eq:14d}
\end{align}
\end{subequations}
These operators satisfy the $\mathscr{N} = 2$ supersymmetry algebra with the trivial central extension
\begin{subequations}
\begin{align}
&((-1)^{F})^{2} = 1, \label{eq:15a}\\
&(Q^{\pm})^{2} = 0, \label{eq:15b}\\
&Q^{\pm}(-1)^{F} = -(-1)^{F}Q^{\pm}, \label{eq:15c}\\
&Q^{+}Q^{-} + Q^{-}Q^{+} = H + c, \label{eq:15d}
\end{align}
\end{subequations}
where $c = 1/L(\alpha)^{2}$ is the trivial center that commutes with all the operators.
In a generic basis, these operators that act on the Hilbert space $\mathcal{H} = L^{2}(0,\ell) \otimes \mathbb{C}^{2}$ are given by $H = h \otimes 1_{2}$, $(-1)^{F} = 1 \otimes Z$, $Q^{\pm} = A_{\alpha}^{\pm} \otimes XP_{\pm}$, where $X$ is a generic $2 \times 2$ hermitian unitary matrix that satisfies the anticommutation relation $XZ = -XZ$ and $XP_{\pm}X = P_{\mp}$.

It should be noted that, under the supersymmetric boundary conditions, $Q^{+}$ and $Q^{-}$ are indeed adjoint to each other such that the combination $H + c$ becomes positive semidefinite.
In other words, the spectrum of the Hamiltonian $\mathrm{Spec}(H)$ is bounded from below, $\mathrm{Spec}(H) \geq -c = -1/L(\alpha)^{2}$, and the lower bound is saturated by the zero-modes of $Q^{\pm}$.

%-----------------------------------------------------------------------------------------------
% SECTION 2.4
%-----------------------------------------------------------------------------------------------
\subsection{$\mathscr{N}=2$ superspectrum}
In order to compute Berry's connection explicitly we have to know the spectrum and normalized energy eigenfunctions $\{\bm{\Psi}_{\pm,n} = \Psi_{\pm,n}\bm{e}_{\pm}\}_{n=0}^{\infty}$.
It is a straightforward exercise to solve the Schr\"{o}dinger equations $h\Psi_{\pm,n} = E_{n}\Psi_{\pm,n}$ for the components $\Psi_{\pm}$ with the supersymmetric boundary conditions in table \ref{tab:1}.
Here are the solutions:
\begin{enumerate}
\item \textit{Type DD.}
In this case the positive energy levels are all doubly degenerate and normalized energy eigenfunctions are found to be of the forms
\begin{subequations}
\begin{align}
\Psi_{+,n}(x)
&= 	\sqrt{\frac{2}{\ell}}\sin\left(\sqrt{E_{n}}x\right), \label{eq:16a}\\
\Psi_{-,n}(x)
&= 	\sqrt{\frac{2}{\ell}\frac{1}{1 + E_{n}L(\alpha)^{2}}}
	\left[\sin\left(\sqrt{E_{n}}x\right) + \sqrt{E_{n}}L(\alpha)\cos\left(\sqrt{E_{n}}x\right)\right], \label{eq:16b}
\end{align}
\end{subequations}
where $E_{n} = (n\pi/\ell)^{2}$, $n = 1,2,3,\cdots$.
In addition to these positive energy eigenstates, there exists a single negative energy eigenstate at $E = E_{0} = -1/L(\alpha)^{2}$ in the ``fermionic'' sector, which is the ground state of the model and takes the following form:
\begin{align}
\Psi_{-,0}(x)
= 	\sqrt{\frac{2}{L(\alpha)}\frac{1}{\mathrm{e}^{2\ell/L(\alpha)} - 1}}
	\exp\left(\frac{x}{L(\alpha)}\right). \label{eq:17}
\end{align}

\item \textit{Type RR.}
As in the previous case the positive energy levels are doubly degenerate too and normalized energy eigenfunctions take the following forms:
\begin{subequations}
\begin{align}
\Psi_{+,n}(x)
&= 	\sqrt{\frac{2}{\ell}\frac{1}{1 + E_{n}L(\alpha)^{2}}}
	\left[\sin\left(\sqrt{E_{n}}x\right) - \sqrt{E_{n}}L(\alpha)\cos\left(\sqrt{E_{n}}x\right)\right], \label{eq:18a}\\
\Psi_{-,n}(x)
&= 	\sqrt{\frac{2}{\ell}}\sin\left(\sqrt{E_{n}}x\right), \label{eq:18b}
\end{align}
\end{subequations}
where $E_{n} = (n\pi/\ell)^{2}$, $n=1,2,3,\cdots$.
In addition, there is a single negative energy eigenstate at $E = E_{0} = -1/L(\alpha)^{2}$ in the ``bosonic'' sector, whose eigenfunction is given by
\begin{align}
\Psi_{+,0}(x)
= 	\sqrt{\frac{2}{L(\alpha)}\frac{1}{1 - \mathrm{e}^{-2\ell/L(\alpha)}}}
	\exp\left(-\frac{x}{L(\alpha)}\right). \label{eq:19}
\end{align}

\item \textit{Type DR.}
In this case all the energy levels are doubly degenerate and normalized energy eigenfunctions take the following forms:
\begin{subequations}
\begin{align}
\Psi_{+,n}(x)
&= 	\sqrt{\frac{2}{\ell + \frac{L(\alpha)}{1 + E_{n}L(\alpha)^{2}}}}
	\sin\left(\sqrt{E_{n}}x\right), \label{eq:20a}\\
\Psi_{-,n}(x)
&= 	\sqrt{\frac{2}{\ell + \frac{L(\alpha)}{1 + E_{n}L(\alpha)^{2}}}}
	\sin\left(\sqrt{E_{n}}(\ell - x)\right), \label{eq:20b}
\end{align}
\end{subequations}
where $E_{n}$ ($E_{0} < E_{1} < \cdots$) are given by the real roots of the transcendental equation
\begin{align}
\tan\left(\sqrt{E}\ell\right) = -\sqrt{E}L(\alpha). \label{eq:21}
\end{align}
We note that the ground state energy $E_{0}$ becomes negative when $-\ell < L(\alpha) < 0$.

\item \textit{Type RD.}
As in the previous case all the energy levels are doubly degenerate too and normalized energy eigenfunctions take the following forms:
\begin{subequations}
\begin{align}
\Psi_{+,n}(x)
&= 	\sqrt{\frac{2}{\ell - \frac{L(\alpha)}{1 + E_{n}L(\alpha)^{2}}}}
	\sin\left(\sqrt{E_{n}}(\ell - x)\right), \label{eq:22a}\\
\Psi_{-,n}(x)
&= 	\sqrt{\frac{2}{\ell - \frac{L(\alpha)}{1 + E_{n}L(\alpha)^{2}}}}
	\sin\left(\sqrt{E_{n}}x\right), \label{eq:22b}
\end{align}
\end{subequations}
where $E_{n}$ ($E_{0} < E_{1} < \cdots$) are the real roots of the transcendental equation
\begin{align}
\tan\left(\sqrt{E}\ell\right) = \sqrt{E}L(\alpha). \label{eq:23}
\end{align}
\end{enumerate}

%-----------------------------------------------------------------------------------------------
% SECTION 3
%-----------------------------------------------------------------------------------------------
\section{Non-Abelian Berry's connection: The Wu-Yang-like monopole} \label{sec:3}
Let us finally study geometric phase in $\mathscr{N} = 2$ supersymmetric quantum mechanics in this system, whose parameter space is $\mathcal{M}_{\text{SUSY}} = U(1) \times U(2)/(U(1) \times U(1))$.
To this end, let us consider a simple time-dependent situation in which the $U(1) \cong S^{1}$ parameter is kept fixed yet the coset $U(2)/(U(1) \times U(1)) \cong S^{2}$ parameters are adiabatically driven along a closed loop $\gamma$ on $S^{2}$.
Then, if the initial state is an element of the subspace $\mathcal{H}_{n} = \mathrm{span}\{\bm{\Psi}_{+,n}, \bm{\Psi}_{-,n}\}$, the final state remains in the subspace $\mathcal{H}_{n}$ and coincides with the initial states up to the time-dependent trivial dynamical phase $\exp(-iE_{n}T)$ and the time-independent nontrivial non-Abelian geometric phase $W_{\gamma}(A)$ given by the formula \cite{Wilczek:1984dh}:
\begin{align}
W_{\gamma}(A)
= 	\mathcal{P}\exp\left(i\oint_{\gamma}A\right), \label{eq:24}
\end{align}
where $A = \left(\begin{smallmatrix}A_{++}&A_{+-}\\ A_{-+}&A_{--}\end{smallmatrix}\right)$ is the hermitian matrix-valued 1-from on the parameter space $U(2)/(U(1) \times U(1)) \cong S^{2}$ defined by
\begin{align}
A_{\alpha\beta}
= 	i\int_{0}^{\ell}\!\!dx\,
	\bm{\Psi}_{\alpha,n}^{\dagger}(x)d\bm{\Psi}_{\beta,n}(x),
	\qquad \alpha,\beta \in \{+,-\}. \label{eq:25}
\end{align}
Here $d$ stands for the exterior derivative on $S^{2}$.
Substituting the solutions one readily finds that the Berry connection takes the following form:
\begin{align}
A
= 	\begin{pmatrix}
	i\bm{e}_{+}^{\dagger}d\bm{e}_{+} 			& iK_{n}(\alpha)\bm{e}_{+}^{\dagger}d\bm{e}_{-} \\
	iK_{n}(\alpha)\bm{e}_{-}^{\dagger}d\bm{e}_{+} 	& i\bm{e}_{-}^{\dagger}d\bm{e}_{-}
	\end{pmatrix}, \label{eq:26}
\end{align}
where $K_{n}(\alpha)$ is the overlap integral between the components
\begin{align}
K_{n}(\alpha)
= 	\int_{0}^{\ell}\!\!dx\,
	\Psi_{+,n}(x)\Psi_{-,n}(x). \label{eq:27}
\end{align}
Under the local gauge transformation $\bm{\Psi}_{\alpha,n} \mapsto \Tilde{\bm{\Psi}}_{\alpha,n} = \bm{\Psi}_{\beta,n}g_{\beta\alpha}$, where $g = (g_{\beta\alpha})$ is a $2 \times 2$ unitary matrix that depends on the $S^{2}$ parameters, the Berry connection transforms as follows:
\begin{align}
A
\mapsto \Tilde{A}
= 	g^{\dagger}Ag + ig^{\dagger}dg
= 	g^{\dagger}Ag - i(dg^{\dagger})g. \label{eq:28}
\end{align}
For the following discussions it is convenient to work in the gauge given by
\begin{align}
g
= 	\begin{pmatrix}
	\bm{e}_{+}^{\dagger} \\
	\bm{e}_{-}^{\dagger}
	\end{pmatrix}. \label{eq:29}
\end{align}
In this gauge the Berry connection takes the simple form $\Tilde{A} = \frac{i}{2}(1 - K_{n}(\alpha))ZdZ$ \cite{Ohya:2014ska}.
Let us now parameterize the hermitian unitary matrix $Z$ as $Z = \frac{\bm{r}}{r} \cdot \bm{\sigma}$, where $\bm{r} = (x_{1}, x_{2}, x_{3}) \in \mathbb{R}^{3}\setminus\{\bm{0}\}$ and $r = \sqrt{x_{1}^{2} + x_{2}^{2} + x_{3}^{2}}$.
A straightforward calculation gives $\Tilde{A} = \Tilde{A}_{i}dx_{i}$, where
\begin{align}
\Tilde{A}_{i}
= 	\epsilon_{ijk}\frac{x_{j}}{r^{2}}\frac{\sigma_{k}}{2}(1 - K_{n}(\alpha)). \label{eq:30}
\end{align}
Notice that, when $K_{n}(\alpha) = 0$, this is nothing the Wu-Yang magnetic monopole which is a classical solution of pure $SU(2)$ Yang-Mills gauge theory.
(For the Wu-Yang monopole, see e.g. \cite{Actor:1979in} for review.)
Note also that, as shown in \cite{Ohya:2014ska}, \eqref{eq:30} is gauge equivalent to the non-Abelian Berry connection discussed in the context of geometric phase in the adiabatic decoupling limit of diatomic molecule \cite{Moody:1985ty}. (See also \cite{Zee:1988,Chandrasekharan:2006wn}.)

%-----------------------------------------------------------------------------------------------
% SECTION 4
%-----------------------------------------------------------------------------------------------
\section{Conclusions and discussions} \label{sec:4}
In this paper we discussed non-Abelian geometric phase in $\mathscr{N} = 2$ supersymmetric quantum mechanics with point-like interactions.
We computed Berry's connection on the parameter space of supersymmetric point-like interactions and showed that it is given by the Wu-Yang-like magnetic monopole in $SU(2)$ Yang-Mills gauge theory.
It would be interesting to point out here that, if $K_{n}$ was a function of $r$ and had the asymptotic behaviors $K_{n}(r) \to 1$ as $r \to 0$ and $K_{n}(r) \to 0$ as $r \to \infty$, the Berry connection would be the celebrated singularity-free 't Hooft-Polyakov monopole \cite{'tHooft:1974qc,Polyakov:1974ek} in $SU(2)$ Yang-Mills-Higgs theory.
It would be also quite interesting to point out that, if the zero modes of $Q^{+}$ and $Q^{-}$ coexisted in the spectrum, the overlap integral $K_{0}(\alpha) = \int_{0}^{\ell}dx\,\Psi_{+,0}(x)\Psi_{-,0}(x)$ between the components \eqref{eq:17} and \eqref{eq:19} would become
\begin{align}
K_{0}(\alpha)
= 	\frac{\ell/L(\alpha)}{\sinh(\ell/L(\alpha))}. \label{eq:31}
\end{align}
The Berry connection \eqref{eq:30} would then take the form
\begin{align}
\Tilde{A}_{i}
= 	\epsilon_{ijk}\frac{x_{j}}{r^{2}}\frac{\sigma_{k}}{2}
	\left(
	1 - \frac{\ell/L(\alpha)}{\sinh(\ell/L(\alpha))}
	\right), \label{eq:32}
\end{align}
which exactly coincides with the Bogomolny-Prasad-Sommerfield (BPS) monopole \cite{Prasad:1975kr,Bogomolny:1975de} in $SU(2)$ Yang-Mills-Higgs theory under the identification $\ell/L(\alpha) \equiv evr$, where $e$ is the electric charge and $v$ is the vacuum expectation value of Higgs field.
As in the model studied in \cite{Sonner:2008be}, it might be interesting to explore a model of supersymmetric point-like interactions in which the ``bosonic'' and ``fermionic'' zero-modes coexist and the Berry connection is given by the 't Hooft-Polyakov monopole that saturates the BPS bound.

%-----------------------------------------------------------------------------------------------
% ACKNOWLEDGMENTS
%-----------------------------------------------------------------------------------------------
\ack
This work is supported in part by ESF grant CZ.1.07/2.3.00/30.0034.

%-----------------------------------------------------------------------------------------------
% REFERENCES
%-----------------------------------------------------------------------------------------------

\end{document}